\documentclass[reprint, superscriptaddress, breaklinks=true, showkeys, showpacs=false, nofootinbib]{revtex4}
\usepackage[T1]{fontenc}
\usepackage[utf8]{inputenc}
\usepackage{comment}
\usepackage{color}
\usepackage{babel}
\usepackage{float}
\usepackage{amssymb}
\usepackage{graphicx}
\usepackage{amsmath}
\usepackage{mathtools}
\usepackage{subfigure}
\usepackage[unicode=true,pdfusetitle,
 bookmarks=true,bookmarksnumbered=false,bookmarksopen=false,
 breaklinks=true,pdfborder={0 0 0},backref=false,colorlinks=true]
 {hyperref}
\usepackage{breakurl}

\makeatletter
\@ifundefined{textcolor}{}
{%
 \definecolor{BLACK}{gray}{0}
 \definecolor{WHITE}{gray}{1}
 \definecolor{RED}{rgb}{1,0,0}
 \definecolor{GREEN}{rgb}{0,1,0}
 \definecolor{BLUE}{rgb}{0,0,1}
 \definecolor{CYAN}{cmyk}{1,0,0,0}
 \definecolor{MAGENTA}{cmyk}{0,1,0,0}
 \definecolor{YELLOW}{cmyk}{0,0,1,0}
}

\hypersetup{colorlinks=true,citecolor=blue,linkcolor=cyan,urlcolor=blue,filecolor= green, breaklinks=true}
\usepackage{url}
\usepackage{breakurl}

\makeatother

\begin{document}
\title{Aspects of quantum states asymmetry for the magnetic dipolar interaction dynamics}

\author{Douglas F. Pinto}
    \email[Electronic address: ]{douglasfpinto@gmail.com}
\author{Jonas Maziero}   
    \email[Electronic address: ]{jonas.maziero@ufsm.br}  
    
    \affiliation{Departamento de F\'isica, Centro de Ci\^encias Naturais e Exatas, Universidade Federal de Santa Maria, Avenida Roraima 1000, 97105-900, Santa Maria, RS, Brazil}

\date{\today} 

\begin{abstract}
We investigate the asymmetry properties of quantum states in relation to the Hamiltonian responsible for the magnetic dipolar interaction (MDI) dynamics, and we evaluate its relationship to entanglement production. We consider some classes of pure and mixed quantum states of two qubits evolved under MDI and, using the asymmetry measure defined via the Wigner-Yanase skew information, we describe the asymmetry dependence on the Hamiltonian parameters and initial conditions of the system. In addition, we define and calculate the dynamics of the asymmetry of local states, characterizing their temporal and interaction parameters dependence. 
Finally, because the MDI Hamiltonian has a null eigenvalue, the group generator-based asymmetry measure does not adequately quantify the state susceptibility with respect to the action of the subspace generated by the eigenvectors associated with this eigenvalue. For this reason, we also define and study the group element-based asymmetry measure with relation to the unitary operator associated with the MDI Hamiltonian.
\end{abstract}

\keywords{Quantum asymmetry, Quantum entanglement, Quantum coherence, Magnetic dipolar interaction}

\maketitle

\section{Introduction}
Research on the intrinsically quantum properties present in physical systems allows the gaining of a better understanding of the fundamental characteristics of nature and has promoted the construction of methods that contribute to the description, transmission, and manipulation of information contained in its microscopic compounds, leading to developments that have a positive impact on diverse areas of Science \cite{Ac_n_2018,Adesso_LoFranco_Parigi2018}. Such properties are primordial for the execution of certain tasks that are impracticable in the context of classical physics. For instance, entanglement is an indispensable attribute in the composition of quantum communication protocols such as teleportation \cite{Bennett1993}, a promising mechanism for the development of a future quantum internet \cite{Pirandola_Braunstein2016}, in the realization of controlled gates applicable in quantum computing, and in the implementation of quantum simulations, where well controlled quantum systems can be used to reproduce the behavior of complex uncontrollable systems \cite{Ac_n_2018},  in addition to enabling several other useful applications in the progress of quantum information science technologies \cite{Dowling-Milburn2003,MasoudMohseni2017}. 

Recently, the resource theory of asymmetry has received attention in the area of Quantum Information Science, favoring applications in various contexts involving both single and composite quantum systems \cite{Girolami2017,Marvian2013,MarvianThesis, Marvian2014, Marvian-Spekkens2014, Piani2016, Marvian-Spekkens-Zanardi2016, D.J.Zhang2017, C.Zhang2017, Bu2018, Marvian2019, Takagi2019}. For example, in multipartite systems the states which break the dynamic symmetry associated with additive Hamiltonians make it possible to quantify or to witness entanglement by the Wigner-Yanase skew information \cite{Girolami2017,Wigner1963}, a function that has the properties necessary for characterization of asymmetry measures in the resource theory  of asymmetry \cite{Takagi2019}. 
Actually, the relationships between skew information measures and quantum correlations have been studied for some time now \cite{Luo-Fu-Oh2012}.
And recently, in Ref. \cite{Dai_2020}, the behavior of the non-classicality  of spins systems was studied using the measure of Wigner-Yanase skew information as a quantifier. Using the contextual re-interpretation of state asymmetry in relation to local unitary groups, which are associated with additive Hamiltonian eigenvectors, it was shown that the difference between the asymmetry of a general global state and the asymmetry of the composition of local states is related to total correlations \cite{Li_2020}.

Asymmetry has also been investigated from an open systems perspective, where it is a useful resource in the processing of quantum information, which in turn suffers deterioration due to interactions between system and environment. However, the freezing of asymmetry in scenarios where open systems are restricted by superselection rules have been studied, proving useful mainly in occasions where noise is described by invariant operations, which enables asymmetry conservation mechanisms in the presence of noise described by such operations \cite{D.J.Zhang2017}, an advantage that can be employed in noisy quantum channels. 

Alike to quantum coherence, state asymmetry properties in relation to the generator of a quantum evolution can play important roles in the entanglement produced due to interactions between subsystems \cite{Pinto-Maziero2018}. Moreover, in a more general context, asymmetry properties have useful characteristics in applications such as in: Alignment of reference frames \cite{Bartlett2009,Bartlett2007}, quantum speed limits  \cite{Marvian-Spekkens-Zanardi2016},  quantum metrology \cite{Takagi2019,C.Zhang2017,Banik2017}, quantum thermodynamics \cite{Marvian2019, Lostaglio-Jennings-Rudolph2015, Lostaglio-Matteo2019}, and quantum communication \cite{Bu2018,Duan-Lukin-Cirac-Zoller2001}. The Wigner-Yanase asymmetry, associated with the Hamiltonian, measures how far the state of the system is from sharing the same base of eigenstates with the observable, established by the commutation relationship between the state of the system and the Hamiltonian associated with its dynamics. Therefore, if the quantum system breaks the dynamic symmetry generated by the Hamiltonian, the commutator is different from zero and, consequently, we say that the state of the system presents non-zero asymmetry.


The presence of  the MDI in several physical systems that have properties of coherence and quantum correlations, such as spins systems \cite{Furman2011, Furman2012, Neumann2010}, nitrogen vacancy centers in diamond \cite{Dolde2013, Choi2019}, and rotational states of molecules \cite{Yun2015},  has motivated research in the context of Quantum Information Science in order to analyze the usefulness of the MDI in the application of resources for the execution of quantum gates for quantum computing \cite{Yun2015, Zhang2020}, quantum simulation execution \cite{Zhou2015}, realization of quantum channels for quantum communication, besides stimulating the investigation of quantum properties of Gibbs thermal states \cite{Kuznetsova2013, Furman2014, Castro2016} and the dynamics of quantum correlations and entanglement \cite{Furman2008, Mohamed2013, Hu2015, Khan2016, Namitha2018, Pinto-Maziero2018}. More recently, it has been shown that, for dipole-dipole interaction and two-photon resonance between two qubits and a coherent cavity field, the dipolar interaction can contribute to robustness against intrinsic decoherence and preserve a higher entanglement rate \cite{Mohamed_2020}. In addition, the MDI plays the role of noise source in several physical systems, leading to the decay of the quantum properties of the system \cite{Klauder1962, Annabestani2018, Ota2007}. It has also been shown that for specific configurations of systems interacting via the MDI, they can preserve the entanglement properties and quantum coherence along the dynamics \cite{Pinto-Maziero2018}.

So, it is relevant to investigate the asymmetry properties of the system configurations in relation to MDI Hamiltonian and to analyze the role that asymmetry plays in respect to the entanglement produced along MDI unitary dynamics. In this article, we study the asymmetry of bipartite quantum states  described by two magnetic dipoles that evolve due to the MDI, whose observable associated  with the generator of the unitary dynamics is non additive (non local). As we showed recently \cite{Pinto-Maziero2018}, the existence of local quantum coherence in the initial states (before MDI occurs) is generally sufficient for the production of entanglement during the course of the MDI dynamics. However, we also showed that it is not a necessary condition, since some initial states with null local coherence are also able to produce maximum entanglement. In this particular case, the peculiar property present in the initial global configuration of the system is the anti-alignment of the dipoles, which, in a way, indicates the presence of asymmetry in relation to the Hamiltonian generating the unitary dynamics.


The structure of the remainder of this article is the following. In Sec. \ref{sec:Observables and configurations}, we present the Hamiltonian and the system states that we will evaluate from the point of view of the asymmetry measure associated with the Wigner-Yanase skew information, which is described in Sec. \ref{sec:AWY}. In Sec. \ref{sec:AsymmetryIDM}, we examine the behavior of the asymmetry function in relation to the Hamiltonian of the MDI for pure and mixed product states. In Sec. \ref{sec:local}, we deal with the asymmetry of local states, and, in Sec. \ref{sec:unitary}, we define and study the asymmetry in relation to the unitary evolution operator. We present our conclusions in Sec. \ref{sec:conc}.

\section{Observables and states}
\label{sec:Observables and configurations}
In this section, we describe the dynamics generator, the Hamiltonian, and states we use to study asymmetry subsequently. Let us start by introducing the general shape of the Magnetic Dipolar Interaction (MDI) Hamiltonian, expressed in the same form as in Ref. \cite{Pinto-Maziero2018}:
$\mathcal{H}=D[(\vec{\sigma}\otimes\sigma_{0})\cdot(\sigma_{0}\otimes\vec{\sigma})-3\hat{n}\cdot\vec{\sigma}\otimes\hat{n}\cdot\vec{\sigma}]$,
where $D=\mu_{0}\gamma_{a}\gamma_{b}\hbar^{2}/16\pi r^{3}$ is the parameter that describes the magnitude of magnetic dipole interaction, $r$ is the distance between the dipoles,  $\mu_{0}$ symbolizes the permeability of vacuum,  $\gamma_{a,b}$ represents the gyromagnetic factor of the subsystems $a$ and $b$, respectively, $\vec{\sigma}$ is the vector of Pauli matrices, $\sigma_{0}$ is the identity matrix of dimension $2$ and $\hat{n}\in\mathbb{R}^{3}$ is the unit vector pointing in the direction of the line that connects the dipole centers. Here, we consider Planck's constant  $\hbar=1$ and we fix $D=1$. For simplicity, but without loss of generality, we assume that the dipoles centers lie along the $z$-axis $\left(\hat{n}=\left(0,0,1\right)\right)$. Thus 
\begin{align}
\label{equation:Hdip}
\mathcal{H} & = 2^{-1}(\sigma_{1}\otimes\sigma_{1}+\sigma_{2}\otimes\sigma_{2}-2\sigma_{3}\otimes\sigma_{3})  \nonumber =  0|\Psi_{-}\rangle\langle\Psi_{-}|+2|\Psi_{+}\rangle\langle\Psi_{+}|-(|\Phi_{-}\rangle\langle\Phi_{-}|+|\Phi_{+}\rangle\langle\Phi_{+}|),
\end{align}
where
$|\Psi_{\pm}\rangle=2^{-{1}/{2}}\left(|01\rangle\pm|10\rangle\right)$, 
$|\Phi_{\pm}\rangle=2^{-{1}/{2}}\left(|00\rangle\pm|11\rangle\right)$
form the Bell basis of maximally entangled states, with $\left\{ |0\rangle,|1\rangle\right\}$  being the standard basis, and we use the notation $|xy\rangle\equiv|x\rangle\otimes|y\rangle$ for the tensor product.

We will first consider that the dipoles are prepared in the configuration of pure-product states, so that
\begin{equation}
|\psi_{ab}\rangle\coloneqq|\psi_{a}\rangle\otimes|\psi_{b}\rangle=\left(\alpha_{a}|0\rangle+\beta_{a}|1\rangle\right)\otimes\left(\alpha_{b}|0\rangle+\beta_{b}|1\rangle\right),
\end{equation}
where $\alpha_{a,b}=\cos(\theta_{a,b}/2)$ and $\beta_{a,b}=\sin(\theta_{a,b}/2)$ are the probability amplitudes associated with each particle, and $\theta_{a,b}\in\left[0,2\pi\right]$, i.e, we consider the dipoles' configurations along coaxial rings in the Bloch's sphere representation \cite{Pinto-Maziero2018}. The dynamics of states evolving under the MDI Hamiltonian is given by the unitary operator $U=\exp\left(-i\mathcal{H}t\right)$. For the pure initial states above, ignoring a global phase, the evolved quantum state reads:
\begin{align} 
|\psi_{t}^{ab}\rangle &=e^{-i\mathcal{H}t}|\psi_{ab}\rangle \nonumber \\
& =	\left(\alpha_{a}\beta_{b}\cos t-i\beta_{a}\alpha_{b}\sin t\right)|01\rangle	+  (\beta_{a}\alpha_{b}\cos t-i\alpha_{a}\beta_{b}\sin t)|10\rangle	+  e^{2it}\left(\alpha_{a}\alpha_{b}|00\rangle+\beta_{a}\beta_{b}|11\rangle\right).
\end{align}

In order to evaluate the consequences of increased entropy of the states in the asymmetry relationship with the MDI Hamiltonian, we also consider two classes of mixed states:
\begin{equation}
\rho_{j}^{ab} :=	\rho_{ja}\otimes\rho_{jb}  =	2^{-1}\left(\sigma_{0}+r_{ja}\sigma_{j}\right)\otimes2^{-1}\left(\sigma_{0}+r_{jb}\sigma_{j}\right),
\end{equation}
where $r_{js}=tr\left(\rho_{js}\sigma_{j}\right)\in[-1,1]$ for $s=a,b$. This states correspond to local Bloch's vectors in the $j$ axis for each subsystem, and we shall take $j=1$ or $j=3$. In these cases, the evolution provided by MDI is given by $\rho_{j}^{ab}(t)=U\rho_{j}^{ab} U^{\dagger}$, so that
\begin{align} 
4\rho_{1}^{ab}\left(t\right)= &	\left(1+r_{1a}r_{1b}\right)\left[|\Psi_{+}\rangle\langle\Psi_{+}|+|\Phi_{+}\rangle\langle\Phi_{+}|\right]+\left(1-r_{1a}r_{1b}\right)\left[|\Psi_{-}\rangle\langle\Psi_{-}|+|\Phi_{-}\rangle\langle\Phi_{-}|\right] \nonumber \\
	& +\left(r_{1b}+r_{1a}\right)\left[e^{3it}|\Phi_{+}\rangle\langle\Psi_{+}|+e^{-3it}|\Psi_{+}\rangle\langle\Phi_{+}|\right]+\left(r_{1b}-r_{1a}\right)\left[e^{it}|\Phi_{-}\rangle\langle\Psi_{-}|+e^{-it}|\Psi_{-}\rangle\langle\Phi_{-}|\right]
\end{align}
and
\begin{align}
4\rho_{3}^{ab}\left(t\right)= &	\left(1+r_{3a}\right)\left(1+r_{3b}\right)|00\rangle\langle00|+\left(1-r_{3a}r_{3b}+\left(r_{3a}-r_{3b}\right)\cos\left(2t\right)\right)|01\rangle\langle01| \nonumber \\
	& +i\left(r_{3a}-r_{3b}\right)\sin\left(2t\right)|01\rangle\langle10|-i\left(r_{3a}-r_{3b}\right)\sin\left(2t\right)|10\rangle\langle01| \nonumber \\
	& +\left(1-r_{3a}r_{3b}-\left(r_{3a}-r_{3b}\right)\cos\left(2t\right)\right)|10\rangle\langle10|+\left(1-r_{3a}\right)\left(1-r_{3b}\right)|11\rangle\langle11|.
\end{align}

\section{Wigner-Yanase asymmetry measure}
\label{sec:AWY}
In this section, we present the asymmetry measure based on the skew information of Wigner and Yanase \cite{Wigner1963}. Here we are interested in  quantifying the presence of asymmetry of states in relation to the Hamiltonian that generates the dynamics for MDI.
The Wigner-Yanase asymmetry of an arbitrary state $\rho$ in relation to $\mathcal{H}$ is given by the following expression \cite{Marvian2014}:
\begin{equation} 
\label{equation:Asymetry}
A\left(\rho,\mathcal{H}\right) 
 :=	-\frac{1}{2}tr\left[\rho^{{1}/{2}},\mathcal{H}\right]^{2}   =	tr\left(\rho\mathcal{H}^{2}\right)-tr\left(\rho^{{1}/{2}}\mathcal{H}\rho^{{1}/{2}}\mathcal{H}\right),
\end{equation} 
where $\left[.,.\right]$ represents the commutator. It is easy to verify that the asymmetry of states evolved under a time-independent Hamiltonian is also time-independent, i.e.,
\begin{equation}
A(\rho_{t},\mathcal{H})=A(\rho,\mathcal{H}),
\end{equation}
where $\rho_{t}=U\rho U^{\dagger}$ and $U=e^{-i\mathcal{H}t}$. Besides, for pure states $\sqrt{\rho}=\sqrt{|\psi\rangle\langle\psi|}=|\psi\rangle\langle\psi|$, and the calculation of asymmetry is reduced to computing the variance of the generator of the dynamics:
\begin{equation}
\label{equation:AEP}
A\left(|\psi\rangle\langle\psi|,\mathcal{H}\right)=\langle\psi|\mathcal{H}^{2}|\psi\rangle-\langle\psi|\mathcal{H}|\psi\rangle^{2}.
\end{equation}

Such expressions, besides of being associated to the quantum coherence related to the generator eigenbasis, due to the Hamiltonian's structure considered in this work, they also reveal an important role of their respective eigenvalues, a fact that will become evident in the next sections. Actually, the Hamiltonian of MDI has a null eigenvalue, and the contribution of the subspace associated with such eigenvalue is not captured by the asymmetry measure defined above. This fact will be used as an argument to consider a Wigner-Yanase asymmetry defined using the unitary operation, rather than using the Hamiltonian, for measuring how much a given initial state changes under the action of a given transformation. 

\section{Asymmetry of states in relation to the MDI Hamiltonian}
\label{sec:AsymmetryIDM}

\begin{figure}
\centering
\includegraphics[width=0.66\textwidth]{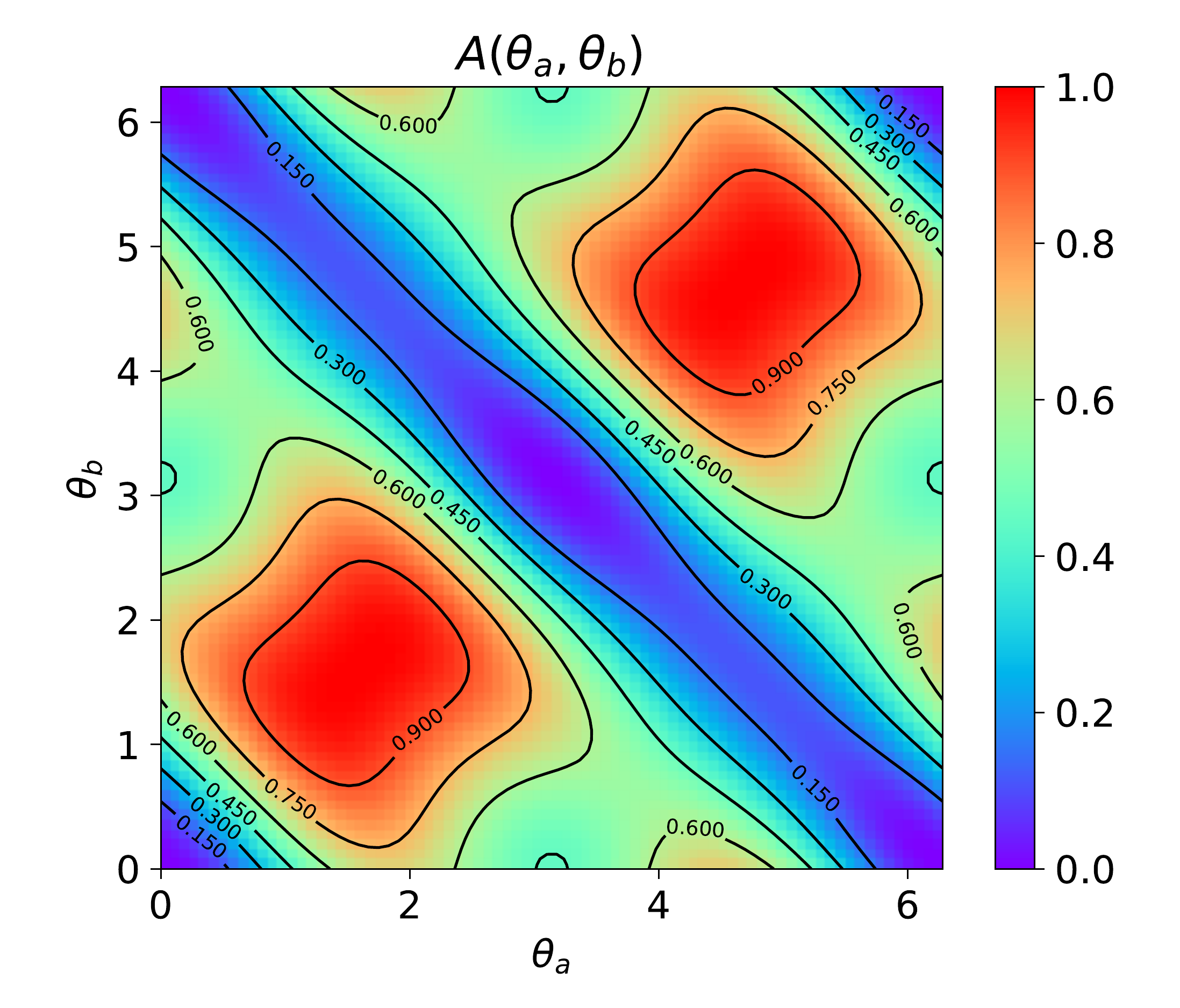}
\caption{\label{fig:Asymmetry} (Color online) Behavior of the Wigner-Yanase asymmetry of states described by $|\psi_{ab}\rangle$ as a function of the initial state parameters $\theta_{a}$ and $\theta_{b}$. It can be seen that the maximum value of the asymmetry function, $A=1$, is reached when the initial state is $|++\rangle$ or $|--\rangle$, and its minimum value, $A=0$, is obtained for the states $|00\rangle$ and $|11\rangle$. Besides, we identify the intermediate values of asymmetry $A=4/9$ and $A=6/9$, which are associated with the initial states $\left\{ |01\rangle,|10\rangle\right\}$  and $\sum_{j=1}^{4}\frac{1}{\sqrt{4}}|E_{j}\rangle$, respectively, with $|E_{j}\rangle$ being the eigenvectors of $\mathcal{H}$.}
\end{figure}

For pure states, using the expression for asymmetry in Eq. (\ref{equation:AEP}), we shall have
\begin{equation}
\label{equation:AsyMDI}
A\left(|\psi_{ab}\rangle,\mathcal{H}\right)=\frac{4}{9}	\left\{ 2\left(\alpha_{a}\beta_{b}+\beta_{a}\alpha_{b}\right)^{2}+\left(\alpha_{a}^{2}\alpha_{b}^{2}+\beta_{a}^{2}\beta_{b}^{2}\right)\right.
	\left.-\left[\left(\alpha_{a}\beta_{b}+\beta_{a}\alpha_{b}\right)^{2}-\left(\alpha_{a}^{2}\alpha_{b}^{2}+\beta_{a}^{2}\beta_{b}^{2}\right)\right]^{2}\right\}.
\end{equation}
The dependency of the asymmetry function on the parameters $\theta_{a}$ and $\theta_{b}$, that define the possible initial state configurations, is illustrated graphically in Fig. \ref{fig:Asymmetry}. We can observe in this figure that the maximum values of asymmetry are reached in the regions around the following two states \begin{equation}|++\rangle=\left(|\Phi_{+}\rangle+|\Psi_{+}\rangle\right)/\sqrt{2}  \text{ and } |--\rangle=\left(|\Phi_{+}\rangle-|\Psi_{+}\rangle\right)/\sqrt{2},
\end{equation}
that are balanced superpositions of a pair of the Hamiltonian eigenstates corresponding to non-zero and distinct eigenvalues. In this case, the maximum asymmetry coincides with the maximum local coherence of each dipole, and corresponds to initial states for which maximum entanglement is generated in some instant of time along the MDI dynamics (see Ref. \cite{Pinto-Maziero2018}). Throughout this article, quantum coherence is measured with relation to the standard basis $\{|0\rangle,|1\rangle\}$.

We observe a similar pattern in the regions of asymmetry values around $6/9$ $(\approx 0,666)$, which are associated with balanced superpositions of all Bell base states. These states correspond to dipole $a$ having null local coherence and dipole $b$ showing maximum coherence, or vice versa:
\begin{align}
& |0\rangle\otimes|\pm\rangle=2^{-1}\left(\pm|\Psi_{+}\rangle\pm|\Psi_{-}\rangle+|\Phi_{+}\rangle+|\Phi_{-}\rangle\right)\text{, } |\pm\rangle\otimes|0\rangle=2^{-1}\left(\pm|\Psi_{+}\rangle\mp|\Psi_{-}\rangle+|\Phi_{+}\rangle+|\Phi_{-}\rangle\right), \\
& |1\rangle\otimes|\pm\rangle=2^{-1}\left(|\Psi_{+}\rangle-|\Psi_{-}\rangle\pm|\Phi_{+}\rangle\mp|\Phi_{-}\rangle\right)\text{, } |\pm\rangle\otimes|1\rangle=2^{-1}\left(|\Psi_{+}\rangle+|\Psi_{-}\rangle\pm|\Phi_{+}\rangle\mp|\Phi_{-}\rangle\right),
\end{align}
and lead to similar intermediate entanglement values ($\approx 0,6$) at specific times ($t=\pi/4$) of the MDI dynamics \cite{Pinto-Maziero2018}. Above, and in what follows, we ignore global phases, once the Wigner-Yanase asymmetry doesn't depend on them.

We notice also the existence of states with asymmetry equal to $4/9$, corresponding to the states 
\begin{equation}
|01\rangle=\frac{1}{\sqrt{2}}\left(|\Psi_{+}\rangle+|\Psi_{-}\rangle\right) \text{ and } |10\rangle=\frac{1}{\sqrt{2}}\left(|\Psi_{+}\rangle-|\Psi_{-}\rangle\right),
\end{equation}
leading to the production of maximum entanglement in some period of the MDI dynamics (as e.g. at $t=\pi/4$). This observation leads us to conclude that the amount of initial state asymmetry does not necessarily establish a direct and unambiguous relationship with entanglement creation.
There are also states with maximum local coherence, such as 
\begin{equation}
|+\rangle\otimes|-\rangle=\frac{1}{\sqrt{2}}\left(|\Phi_{-}\rangle-|\Psi_{-}\rangle\right) \text{ and
} |-\rangle\otimes|+\rangle=\frac{1}{\sqrt{2}}\left(|\Phi_{-}\rangle+|\Psi_{-}\rangle\right),
\end{equation}
that have asymmetry around $1/9$ and also lead to maximum dynamic entanglement. 

From the results presented above, we infer that the presence of asymmetry in relation to the Hamiltonian of the MDI does lead to the existence of entanglement at some time. However, in general there is no direct quantitative relationship between $A$ and $E$. This fact is more evident for initial state configurations involving a pair of states in which one of the elements is the singlet state $|\Psi_{-}\rangle$. This is so because the nullity of the corresponding eigenvalue leads to the non-contribution of this state in the calculation of the asymmetry, reducing its magnitude in these cases. On the other hand, we can observe that the line of states $\theta_{b}=\theta_{a}$ and $\theta_{b}=2\pi-\theta_{a}$ produce null asymmetry in all regions whose configurations are $\theta_{a} = n\pi$ with $n\in\mathbb{Z}$, a behavior analogous to what occurs during the entanglement dynamics \cite{Pinto-Maziero2018}.
So, in the next section, we will calculate the asymmetry of local states, which, in addition to help in understanding the role of local states in the evolution of asymmetry along the dynamics of the MDI, this kind of function also contains the partial contributions of the subspace associated with the null eigenvalue of the Hamiltonian.



\begin{figure}
\centering
\label{fig:first}%
\includegraphics[height=2.8in]{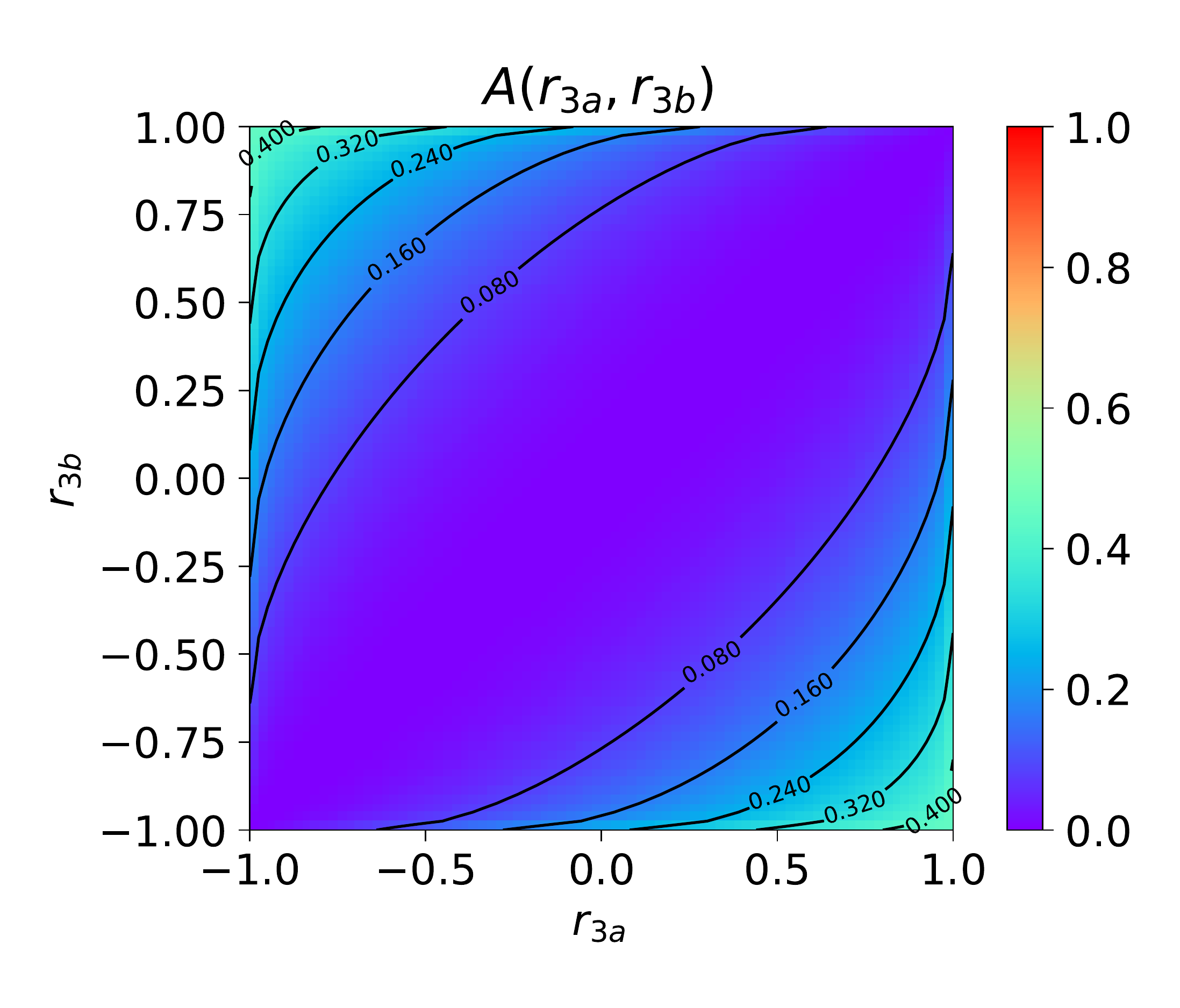}
\qquad
\label{fig:second}%
\includegraphics[height=2.8in]{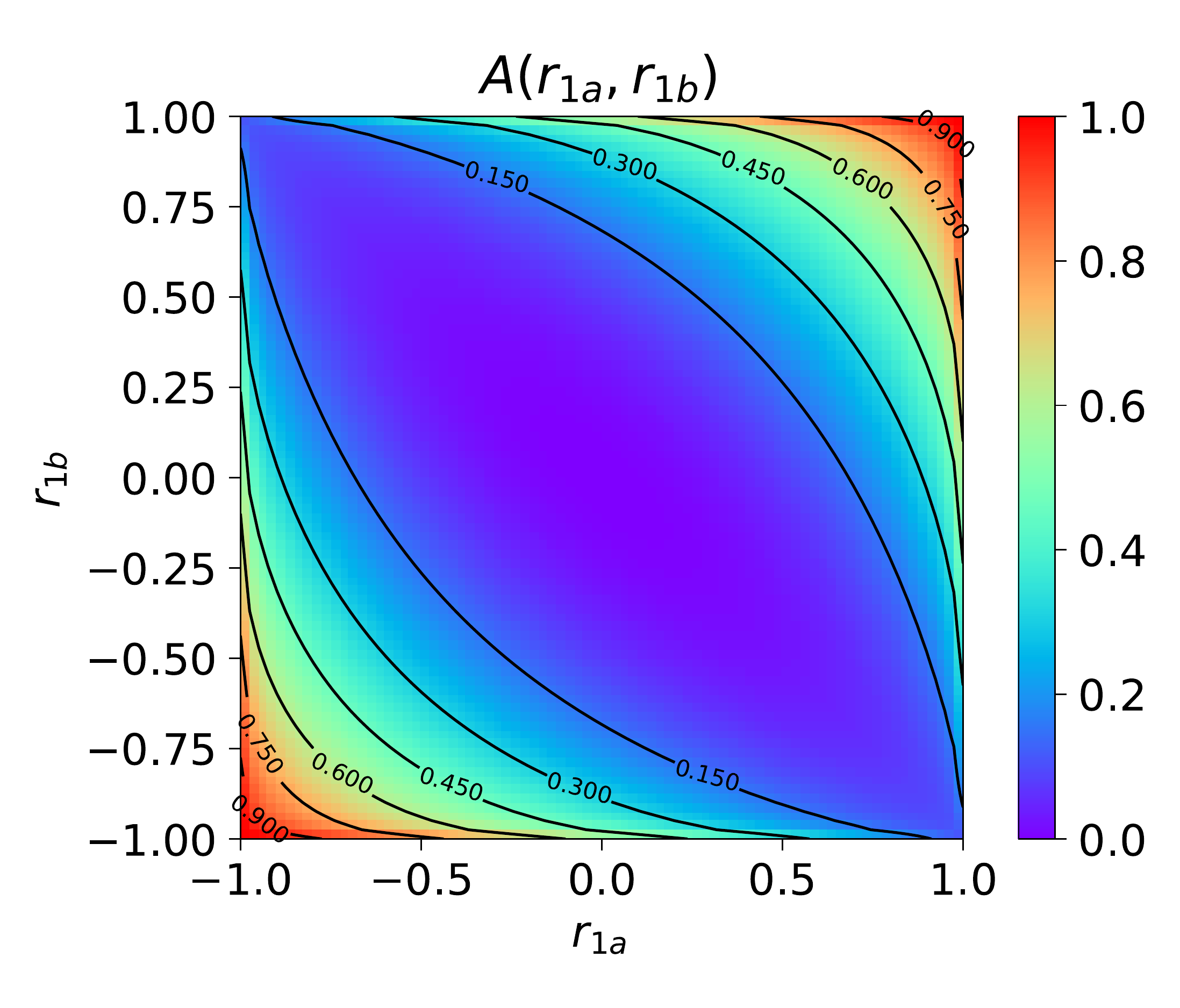}
\caption{(Color online) (a) Wigner-Yanase asymmetry of the $\rho_{3}^{ab}$ class of states, whose quantum coherence is null in the standard basis, as a function of the $r_{3a}$ and $r_{3b}$ parameters. For $r_{3a}=-r_{3b}$ we have $A=4r_{3a}^{2}/9$. So, the maximum value of asymmetry for this class of states is obtained when $r_{3a}=\pm1$ and $r_{3b}=\mp1$, which corresponds to the states $|01\rangle$ and $|10\rangle$, respectively. The line of states where $r_{3a}=r_{3b}$ has asymmetry equal to zero.
(b) Asymmetry of the $\rho_{1}^{ab}$ class of states as a function of the parameters $r_{1a}$ and $r_{1b}$, associated with the Bloch vector of each dipole. For $r_{1a}=r_{1b}$ we have $A=r_{1a}^{2}$. Thus, for such a class of states, the maximum values for asymmetry are obtained for $r_{1a}=\pm1$ and $r_{1b}=\pm1$, which corresponds to the states $|++\rangle$ and $|--\rangle$, respectively. The line of states with $r_{1a}=-r_{1b}$ has $A=r_{1a}^{2}/9$.}
\end{figure}

Regarding the mixed states configurations defined in Sec. \ref{sec:Observables and configurations}, the Wigner-Yanase asymmetry for the class of incoherent states is given by
\begin{equation}
A(\rho_{3}^{ab},\mathcal{H})=2\left(1-r_{3a}r_{3b}-\sqrt{\left(1-r_{3a}^{2}\right)\left(1-r_{3b}^{2}\right)}\right)/9,
 \end{equation}
while for the class of states with coherence the asymmetry reads
\begin{equation}
A\left(\rho_{1}^{ab},\mathcal{H}_{d}\right)=\left(5+4r_{1a}r_{1b}-5\sqrt{\left(1-r_{1a}^{2}\right)\left(1-r_{1b}^{2}\right)}\right)/9.
\end{equation}
These functions are shown in Fig. \ref{fig:second}. These results reveal the main aspects already obtained for pure states, but also show that the decrease in purity leads to the diminishing in the asymmetry of the states.

\section{Asymmetry of local states}
\label{sec:local}

With the study of asymmetry associated with the product state classes developed so far, it was possible to observe that despite positive asymmetry being a necessary condition for the production of entanglement under MDI dynamics, there is no direct relationship between them. Thus, as the asymmetry is not affected by the unitary dynamics of global states, i.e., $A(\rho_{t}^{ab},\mathcal{H})=A(\rho^{ab},\mathcal{H})$, in order to understand the mechanism of temporal evolution of asymmetry and their local contributions, we will evaluate the asymmetry of the local evolved states, i.e., we want to quantify the susceptibility of local states under the action of the dynamics generator $\mathcal{H}$. 
For defining this quantifier, we take the partial trace \cite{ptr} over one of the dipoles, e.g., $$\rho_{t}^{a}=tr_{b}(\rho_{t}^{ab}),$$ and, in order to preserve the correspondence with the Hamiltonian's dimensionality, we compose this reduced state with the maximally mixed state for the other dipole, i.e.,
\begin{equation}
 \tilde{\rho}_{t}^{a} := \rho_{t}^{a}\otimes\frac{\sigma_{0}}{2}.
\end{equation}
Once constructed this state, we defined the local Wigner-Yanase asymmetry of subsystem $A$ as
\begin{equation}
    A_{l}(\rho_{t}^{a},\mathcal{H}):=A\left(\tilde{\rho}_{t}^{a},\mathcal{H}\right)	:=-\frac{1}{2}tr\left[\tilde{\rho}_{t}^{a},\mathcal{H}\right]^{2}.
\end{equation}
The local asymmetry $A_{l}(\rho_{t}^{b},\mathcal{H})$ is defined in a similar way.

\subsection{Local asymmetry for pure-product initial states}
\begin{figure}[H]
\centering
\includegraphics[width=1.04\textwidth]{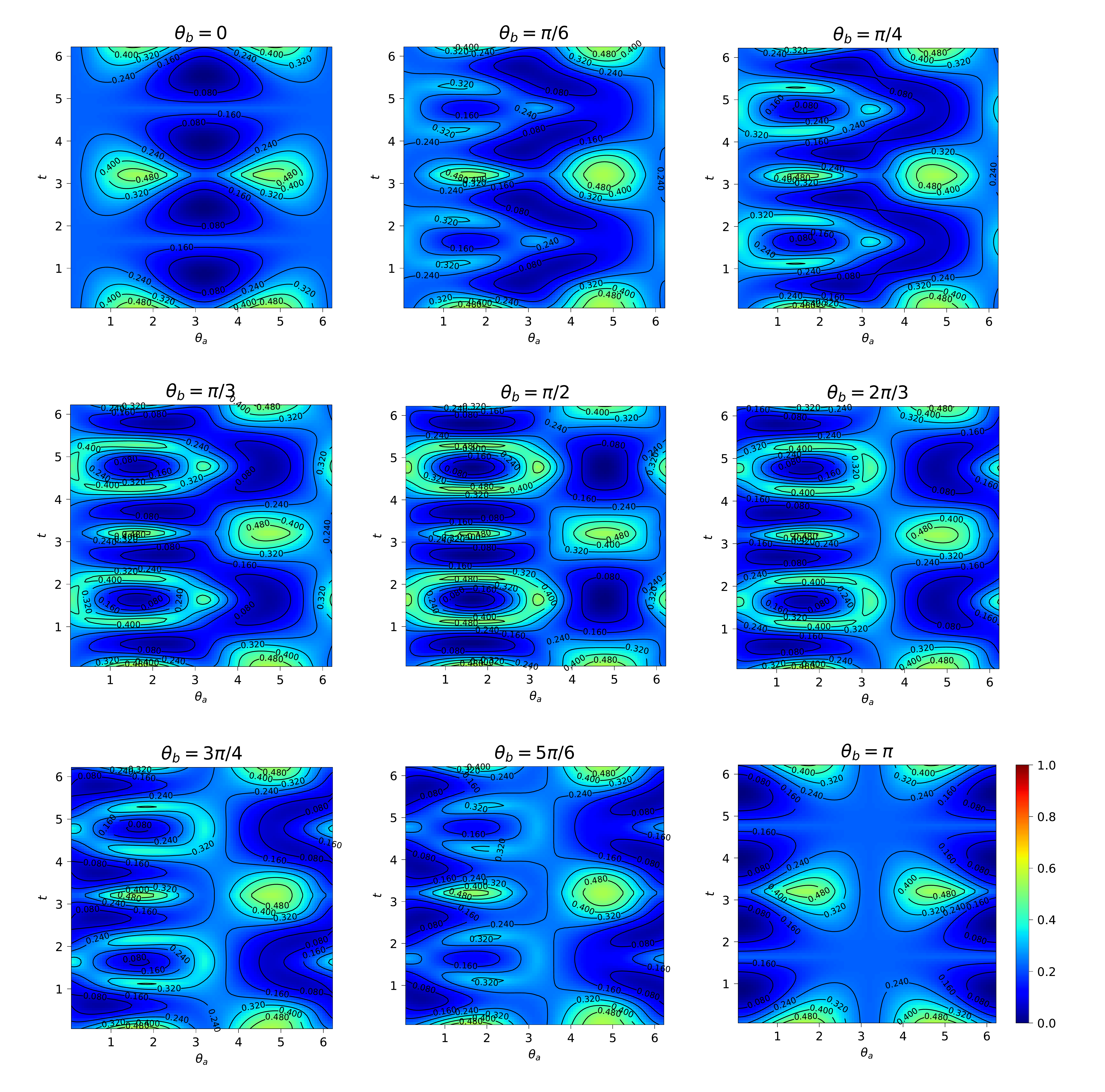}
\caption{\label{fig:LocalAsymmetrythb} (Color online) Dynamics of local asymmetry, $A_{l}(\rho_{t}^{a},\mathcal{H}),$ of dipole $a$ as a function of time and of the pure-product initial states configurations parameters $\theta_{a}$ and $\theta_{b}$. For these plots, we consider the parameter $\theta_{b}$ fixed and allow the parameters $t$ and $\theta_{a}$ to vary ``continuously'' in their respective intervals.}
\end{figure}

For the evolved pure states shown in Sec. \ref{sec:Observables and configurations}, using the state

\begin{align}
\tilde{\rho}_{t}^{a} &=	  tr_{b}|\Psi_{t}^{ab}\rangle\langle\Psi_{t}^{ab}|\otimes\frac{\sigma_{0}}{2} \\
&   =\left\{ \left(\alpha_{a}^{2}\alpha_{b}^{2}+\alpha_{a}^{2}\beta_{b}^{2}\cos^{2}\left(t\right)+\beta_{a}^{2}\alpha_{b}^{2}\sin^{2}\left(t\right)\right)|0\rangle\langle0|+\left(\beta_{a}^{2}\beta_{b}^{2}+\beta_{a}^{2}\alpha_{b}^{2}\cos^{2}\left(t\right)+\alpha_{a}^{2}\beta_{b}^{2}\sin^{2}\left(t\right)\right)|1\rangle\langle1|\right. \nonumber \\
	& +\left[\alpha_{a}\beta_{a}\cos\left(t\right)\left(\alpha_{b}^{2}\exp\left(2it\right)+\beta_{b}^{2}\exp\left(-2it\right)\right)-i\alpha_{b}\beta_{b}\sin\left(t\right)\left(\alpha_{a}^{2}\exp\left(2it\right)+\beta_{a}^{2}\exp\left(-2it\right)\right)\right]|0\rangle\langle1| \nonumber \\
	& \left. +\left[\alpha_{a}\beta_{a}\cos\left(t\right)\left(\beta_{b}^{2}\exp\left(2it\right)+\alpha_{b}^{2}\exp\left(-2it\right)\right)+i\alpha_{b}\beta_{b}\sin\left(t\right)\left(\beta_{a}^{2}\exp\left(2it\right)+\alpha_{a}^{2}\exp\left(-2it\right)\right)\right]|1\rangle\langle0|\right\} \otimes\frac{\sigma_{0}}{2}
\end{align}

we compute the local asymmetry of subsystem $a$. The expression for $A_{l}(\rho_{t}^{a},\mathcal{H})$ is too cumbersome to be shown here. So, this local asymmetry is shown graphically in Fig. \ref{fig:LocalAsymmetrythb} as a function of dipole $a$ initial state and time for some initial states of dipole $b$. From these plots, it is possible to determine the favorable time bands for obtaining the highest values of local asymmetry, that occurs for $t=k\pi$, with $k=0,1,2,\cdots$, and for regions where $\theta_{a}=\pi/2$ or $\theta_{a}=3\pi/2$, regardless of $\theta_{b}$, and for $t=k\pi/3$ in the region where $\theta_{a}=\theta_{b}=\pi/2$. Besides, we can see that $A_{l}(\rho_{t}^{a},\mathcal{H})$ has a period equal to $\pi.$

\begin{figure}[H]
\centering
\includegraphics[width=1.03\textwidth]{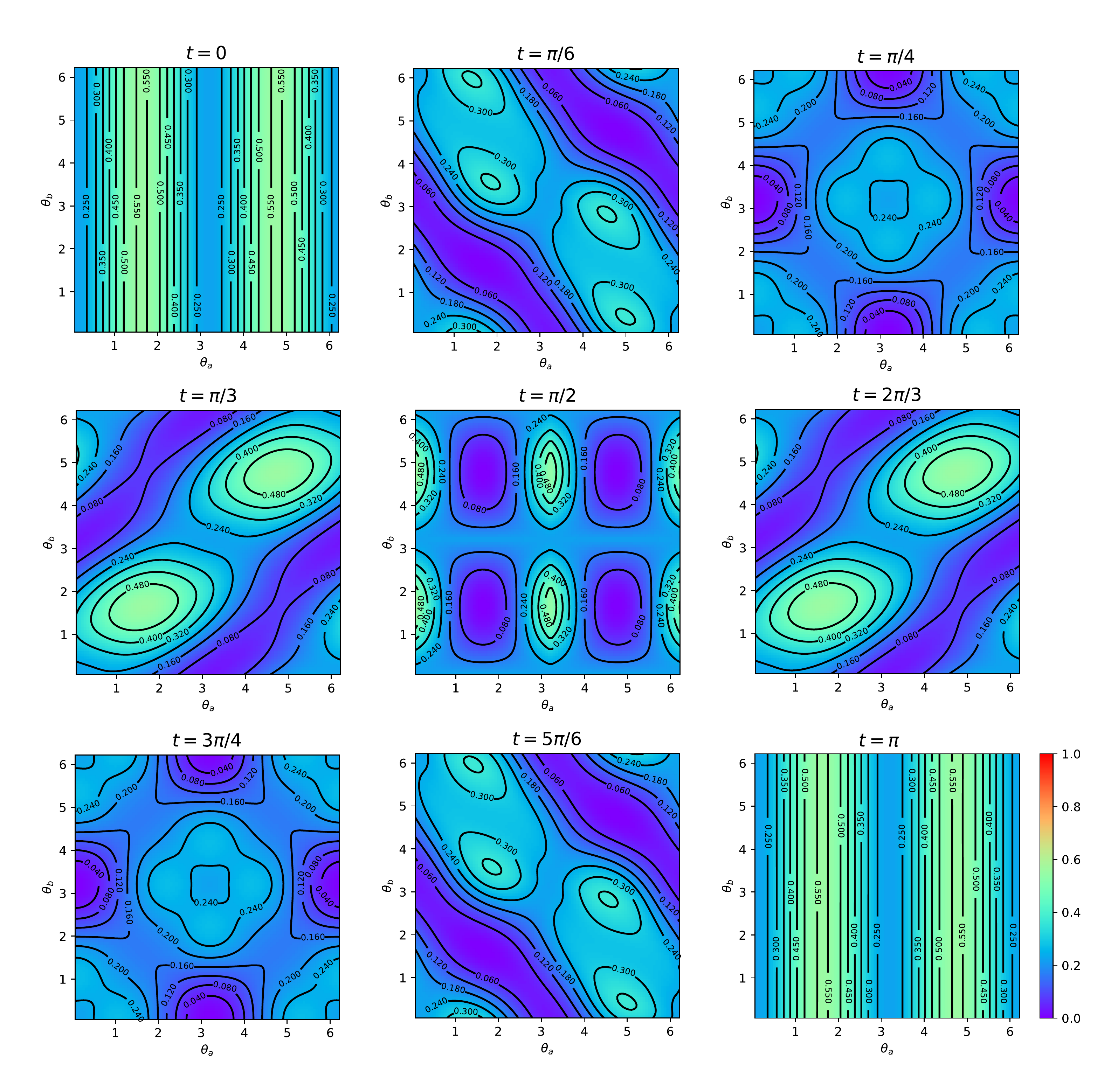}
\caption{\label{fig:LocalAsymmetry} (Color online)
These plots illustrate the behavior of the local asymmetry of subsystem $a$, $A_{l}(\rho_{t}^{a},\mathcal{H}),$ as a function of time and of the parameters of the initial pure-product state configurations represented by $\theta_{a}$ and $\theta_{b}$.}
\end{figure}

In Fig. \ref{fig:LocalAsymmetry}, the local asymmetry is shown as a function of the initial state parameters for some instants of time. The maximum values of $A_{l}(\rho_{t}^{a},\mathcal{H})$ are around $0.55$, and are obtained in the regions of $\theta_{a}=\pi/2$ or $\theta_{a}=3\pi/2$ with $t=k\pi$ for $k = 0,1,2, \cdots.$  
In addition, for $t=\pi/3 $ or $ t=2\pi/3$, in the regions $\theta_{a}=\theta_{b}=\pi/2$ and $ \theta_{a} = \theta_{b}=3\pi/2,$ respectively, the value $ 0.55 $ is also obtained for $A_{l}(\rho_{t}^{a},\mathcal{H})$.

We observe that the general dependence of local asymmetry with the initial state and time is intricate and quite similar with the dependence of local quantum coherence given by the $l_{1}$-norm coherence. Besides, there is even less direct relationship between local asymmetry and entanglement than what was observed for global asymmetry. So, in the next section, we shall study the system state global asymmetry with regard to the element of the group of time transformations, i.e., we shall look at the quantum state susceptibility in relation to the time evolution operator associated with the MDI Hamiltonian.  

\subsection{Local asymmetry for mixed-product initial states}

\begin{figure}[H]
\centering
\includegraphics[width=1.02\textwidth]{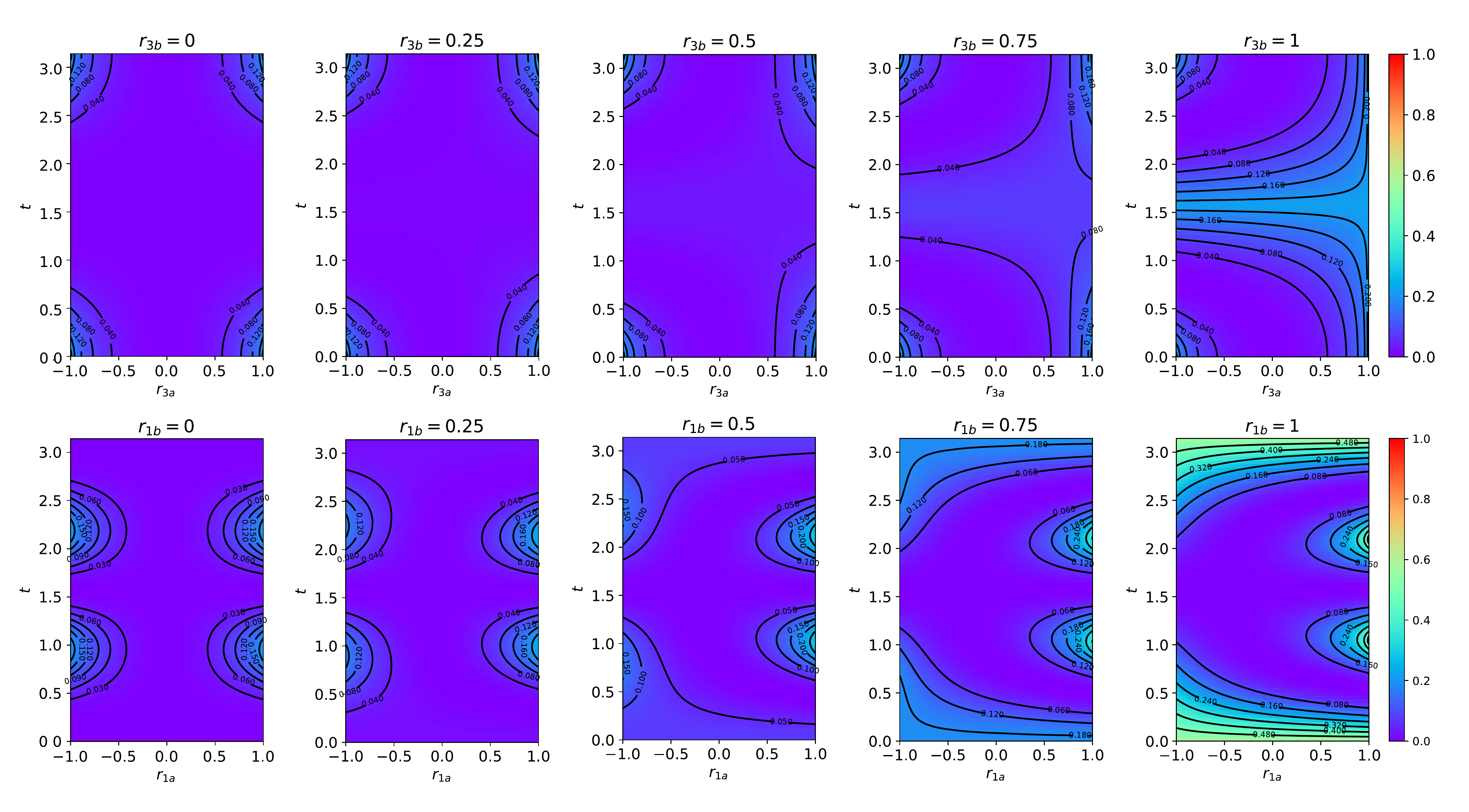}
\caption{\label{fig:rho3LocalAsymmetry} (Color online) Local asymmetry of dipole $a$, $A_{l}(\rho_{j}^{a}(t),\mathcal{H})$, as a function of time and $r_{ja}$ for some values of $r_{jb}$, for the initial states $\rho_{j}^{ab}=\rho_{j}^{a}\otimes\rho_{j}^{b}$ evolved under the MDI Hamiltonian. We see that the local asymmetry has period equal to $\pi$.}
\end{figure}

For the evolved mixed states shown in Sec. \ref{sec:Observables and configurations}, we have the reduced states:
$\rho_{j}^{a}(t)=tr_{b}\left(\rho_{j}^{ab}(t)\right)$. So, the local asymmetry for dipole $a$ is computed using the density matrices:
\begin{align}
    4\tilde{\rho}_{1}^{a}(t)= &	4\rho_{j}^{a}(t) \otimes\frac{\sigma_{0}}{2} \\
	= & \left(\left(r_{1b}+r_{1a}\right)\cos\left(3t\right)-\left(r_{1b}-r_{1a}\right)\cos\left(t\right)\right)\left(|00\rangle\langle10|+|01\rangle\langle11|+|10\rangle\langle00|+|11\rangle\langle01|\right)/8 \nonumber \\
	& + \left(|00\rangle\langle00|+|01\rangle\langle01|+|10\rangle\langle10|+|11\rangle\langle11|\right) \nonumber 
\end{align}
and 
\begin{align}
    \tilde{\rho}_{3}^{a}(t)= &	\rho_{3}^{a}(t)\otimes\sigma_{0}/2 \\
= &	\left(2+\left(r_{3b}+r_{3a}\right)+\left(r_{3a}-r_{3b}\right)\cos\left(2t\right)\right)\left(|00\rangle\langle00|+|01\rangle\langle01|\right)/8 \nonumber \\
	& +\left(2-\left(r_{3b}+r_{3a}\right)-\left(r_{3a}-r_{3b}\right)\cos\left(2t\right)\right)\left(|10\rangle\langle10|+|11\rangle\langle11|\right)/8.
\end{align}
Thereby, the respective asymmetries of these local states are given by the expressions:
\begin{align}
4\sqrt{\frac{A_{l}\left(\rho_{1}^{a}(t),\mathcal{H}\right)}{5}}= & \sqrt{\left(r_{1a}-r_{1b}\right)\cos\left(t\right)+\left(r_{1a}+r_{1b}\right)\cos\left(3t\right)+2} \nonumber \\
&-\sqrt{\left(r_{1b}-r_{1a}\right)\cos\left(t\right)-\left(r_{1a}+r_{1b}\right)\cos\left(3t\right)+2}
\end{align}
and 
\begin{equation}
    3\sqrt{A_{l}\left(\rho_{3}^{a}\left(t\right),\mathcal{H}\right)}=\sqrt{1-r_{3a}+\left(r_{3a}-r_{3b}\right)\sin^{2}\left(t\right)}-\sqrt{1+r_{3a}+\left(r_{3b}-r_{3a}\right)\sin^{2}\left(t\right)}.
\end{equation}
These functions are shown graphically in Fig. \ref{fig:rho3LocalAsymmetry}. The same observations made in the last subsection for pure-product initial states also hold here.

\section{Asymmetry in relation to the unitary operator}
\label{sec:unitary}

In this section, we study the Wigner-Yanase skew information in relation to the unitary operator generated by the magnetic dipolar interaction Hamiltonian. We do this expecting to obtain a more sensitive measure regarding the temporal evolution of the states and which includes the contribution of the phases corresponding to the eigenvalues of the observable responsible for the dynamics. So, we define the unitary asymmetry as the Wigner-Yanase skew information of a state $\rho$ with respect to the non-Hermitian unitary time evolution operator:
\begin{align}
A_{U}\left(\rho,U_{t}\right)& :=	\frac{1}{2}tr\left\{ \left[\sqrt{\rho},U_{t}\right]^{\dagger}\left[\sqrt{\rho},U_{t}\right]\right\} \\
& =	1-tr\left\{ \sqrt{\rho}U_{t}\sqrt{\rho}U_{t}^{\dagger}\right\} ,
\end{align}
For pure states, this function can be recast as
\begin{align}
A_{U}\left(|\psi\rangle,U_{t}\right)	& =1-tr\left\{ |\psi\rangle\langle\psi|U_{t}|\psi\rangle\langle\psi|U_{t}^{\dagger}\right\}     \\
	& =1-|\langle \psi|\psi_{t}\rangle|^{2},
\end{align}
where $|\psi_{t}\rangle=U_{t}|\psi\rangle$. Such a function determines the degree of dissimilarity between the initial prepared state and the state obtained from the evolution dictated by the unitary operator.

To assess the unitary asymmetry in the context of the MDI, we consider the spectral decomposition of the unitary operator given by the following Bell-diagonal matrix:
\begin{equation}
U_{t}=|\Psi_{-}\rangle\langle\Psi_{-}|+\exp\left(-2it\right)|\Psi_{+}\rangle\langle\Psi_{+}|+\exp\left(it\right)\left(|\Phi_{-}\rangle\langle\Phi_{-}|+|\Phi_{+}\rangle\langle\Phi_{+}|\right).
\end{equation}
Considering the same class of pure states of the previous sections, $$|\psi\rangle=|\psi_{ab}\rangle\equiv c_{1}|\Psi_{-}\rangle+c_{2}|\Psi_{+}\rangle+c_{3}|\Phi_{-}\rangle+c_{4}|\Phi_{+}\rangle,$$ with 
\begin{equation}
\begin{cases}
c_{1}=2^{-1/2}\left[\cos\left(\frac{\theta_{a}}{2}\right)\sin\left(\frac{\theta_{b}}{2}\right)-\sin\left(\frac{\theta_{a}}{2}\right)\cos\left(\frac{\theta_{b}}{2}\right)\right],\\
c_{2}=2^{-1/2}\left[\cos\left(\frac{\theta_{a}}{2}\right)\sin\left(\frac{\theta_{b}}{2}\right)+\sin\left(\frac{\theta_{a}}{2}\right)\cos\left(\frac{\theta_{b}}{2}\right)\right],\\
c_{3}=2^{-1/2}\left[\cos\left(\frac{\theta_{a}}{2}\right)\cos\left(\frac{\theta_{b}}{2}\right)-\sin\left(\frac{\theta_{a}}{2}\right)\sin\left(\frac{\theta_{b}}{2}\right)\right],\\
c_{4}=2^{-1/2}\left[\cos\left(\frac{\theta_{a}}{2}\right)\cos\left(\frac{\theta_{b}}{2}\right)+\sin\left(\frac{\theta_{a}}{2}\right)\sin\left(\frac{\theta_{b}}{2}\right)\right],
\end{cases}
\end{equation}
the unitary asymmetry is given by
\begin{equation}
    A_{U}\left(|\psi_{ab}\rangle,U_{t}\right)=1-\left[c_{1}^{2}+c_{2}^{2}\cos\left(2t\right)+\left(c_{3}^{2}+c_{4}^{2}\right)\cos\left(t\right)\right]^{2}-\left[\left(c_{3}^{2}+c_{4}^{2}\right)\sin\left(t\right)-c_{2}^{2}\sin\left(2t\right)\right]^{2}.
\end{equation}

\begin{figure}[H]
\centering
\includegraphics[width=1.03\textwidth]{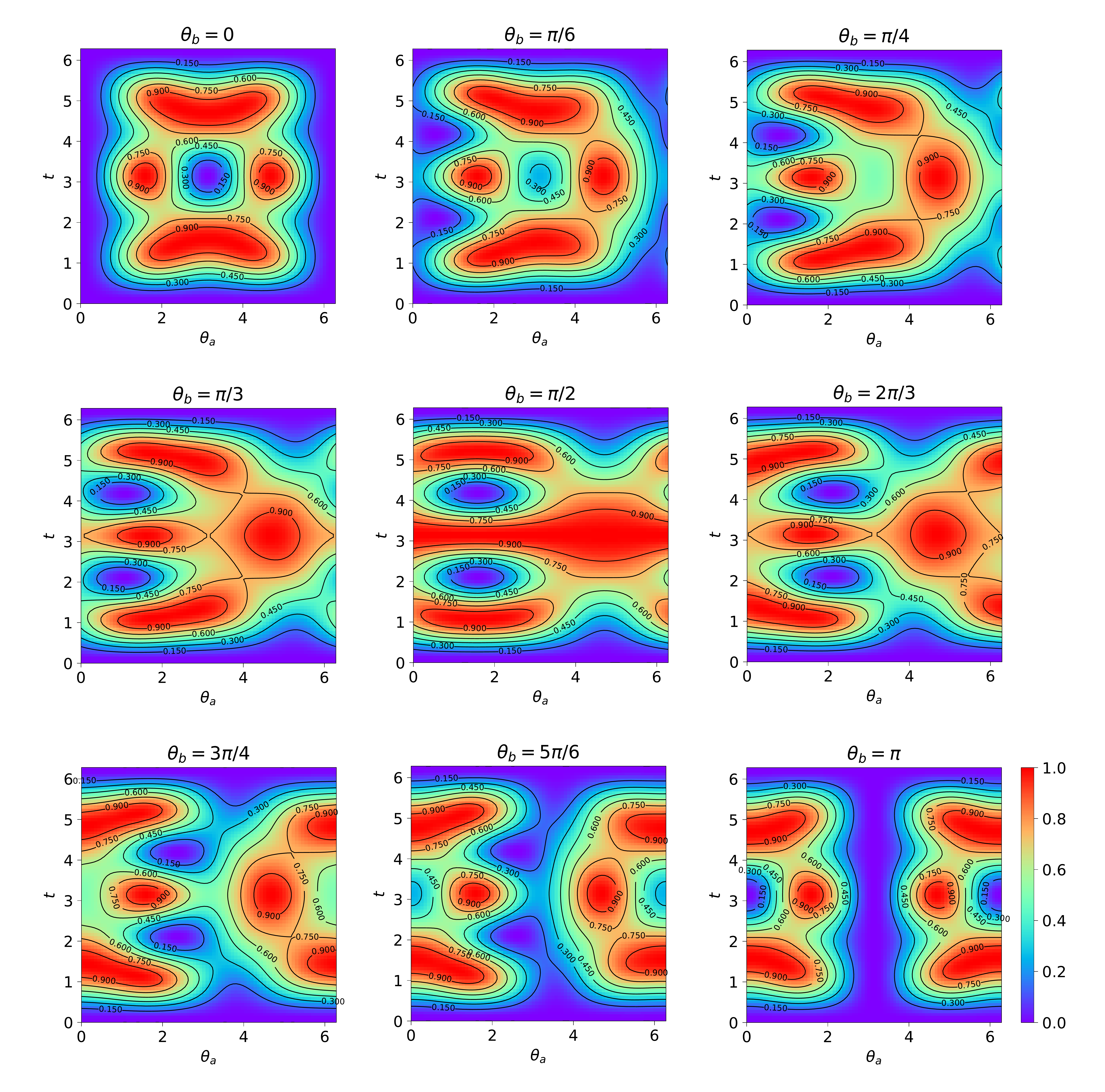}
\caption{\label{fig:UnitSkewInfthb} (Color online)
Behavior of the unitary asymmetry as a function of dipole $a$ initial state parameter $\theta_{a}$ and of the time parameter $t$, for some fixed values of dipole $b$ initial state parameter $\theta_{b}$. This sequence of plots allow us to identify the periods of time resulting in a greater unitary asymmetry $A_{U}$.
}
\end{figure}

\begin{figure}[H]
\centering
\includegraphics[width=1.03\textwidth]{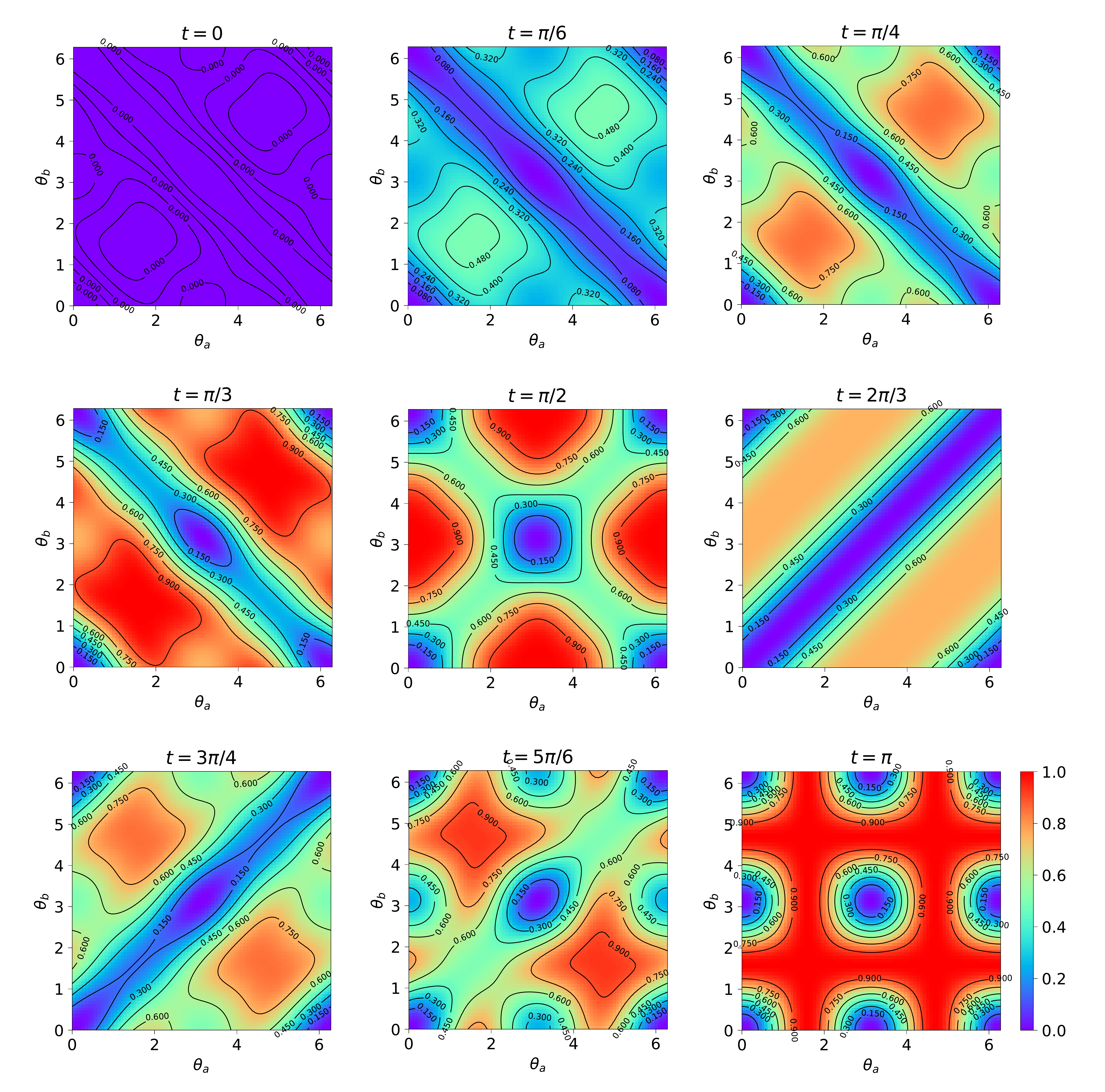}
\caption{\label{fig:UnitSkewInf} (Color online) 
These graphs illustrate the behavior of  asymmetry of states with respect to the MDI unitary operator as a function of the parameters determining the initial conditions, $\theta_{a}$ and $\theta_{b}$, for some values fixed for the time parameter $t$. The only region of states that remain constant throughout the MDI dynamics are those where the parameters $\theta_{a}$ and $\theta_{b}$ corresponds to the states $|00\rangle$ and $|11\rangle$, whose asymmetry $A_{U}$ is null, and which coincides with the regions where the global asymmetry $A\left(|\psi_{ab}\rangle,\mathcal {H}\right)$ is null.
However, in general the state asymmetry $A_{U}$ behaves similarly to the global asymmetry $A\left(|\psi_{ab}\rangle,\mathcal{H}\right)$ when the time parameter is fixed around $\pi/4$. In this figure, we can see that the maximum value of the unitary asymmetry $A_{U}$ is obtained in three periods of the $A$'s dynamics, with regions of different state configurations. For the period $t=\pi/3$, we have maximum asymmetry $A_{U}$ in the region of states with $\theta_{a}=\theta_{b}=\pi/2$ or $\theta_{a}=\theta_{b}=3\pi/2$. For the case where $t =\pi/2$, the maximum unitary asymmetry $A_{U}$ is reached in regions where $\theta_ {a}=\pi$ and $\theta_{b}=0$ or $\theta_{b}=2\pi$ and vice versa. In addition, for the period $t=\pi$, any line of states where $\theta_{a}=\pi/2$ or $\theta_{a}=3\pi/2$ for any regions of $\theta_{b}$, and vice versa, lead to the maximum value of the unitary asymmetry $A_{U}$. }
\end{figure}

Analyzing the behavior of the unitary asymmetry dynamics illustrated in Figs. \ref{fig:UnitSkewInfthb}  and \ref{fig:UnitSkewInf}, and comparing with the evolution of entanglement under the MDI \cite{Pinto-Maziero2018}, we emphasize that the regions of maximum entanglement during the dynamics of the MDI are contained in the regions of states of maximum unitary asymmetry $A_{U}$, although they may occur in different periods of their temporal dynamics. In addition, the regions of null entanglement coincide with regions of null unitary asymmetry. On the other hand, there are also regions with maximum unitary asymmetry but with partial entanglement.

\section{Conclusions}
\label{sec:conc}
In this work, we analyzed the quantum state Wigner-Yanase asymmetry in relation to the Magnetic Dipolar Interaction (MDI) Hamiltonian as the generator of the temporal evolution. We described the dependence of asymmetry in terms of the parameters that define the Hamiltonian and in terms of the initial state  configurations of the established bipartite system, where we considered classes of pure and mixed initial states separable and restricted to real local phases. We obtained analytical expressions for the asymmetry of pure and mixed states, from which it was possible to observe the regions that admit maximum asymmetry and thus establish relations with the purity and entanglement production during the dynamics under the MDI. We also defined the local asymmetry, a quantity that reveals the local states susceptibility under the action of the Hamiltonian generator of the global dynamics under the MDI. Furthermore, in order to quantify the role of Hamiltonian eigenvalues for MDI dynamics, we defined the Wigner-Yanase skew information measure in relation to the MDI unitary operator, obtaining thus a better agreement between states with greater skew-information and states capable of producing entanglement along the MDI dynamics.

\begin{acknowledgments}
This work was supported by the Coordena\c{c}\~ao de Aperfei\c{c}oamento de Pessoal de N\'ivel Superior (CAPES), process 88882.427913/2019-01, and by the Instituto Nacional de Ci\^encia e Tecnologia de Informa\c{c}\~ao Qu\^antica (INCT-IQ), process 465469/2014-0.
\end{acknowledgments}


\end{document}